\documentclass[11pt,a4paper]{article}
\usepackage{jheppub}
\usepackage{bm}  
\usepackage{slashed}
\usepackage{verbatim}
\usepackage{epsf}
\usepackage{latexsym,feynmf}
\usepackage[italian,english]{babel}
\newcommand\be{\begin{equation}}
\newcommand\ee{\end{equation}}
 
\preprint{UCSD-PTH-12-14}

\title{Spontaneous $R$-symmetry breaking from the renormalization group flow}

\author{Antonio Amariti}
\author{and David Stone}

\affiliation{Department of Physics \\
University of California, San Diego \\
La Jolla, CA 92093}

\emailAdd{amariti@ucsd.edu}
\emailAdd{dcstone@physics.ucsd.edu}

\abstract{We propose a mechanism of $R$-symmetry breaking in 
four-dimensional DSB models based on the RG properties of the coupling constants.
By constraining the UV sector, we generate new hierarchies amongst the couplings
that allow a spontaneously 
broken $R$-symmetry in models with pure chiral fields of $R$-charges $R=0$ and $R=2$ only.
The result is obtained by a combination of one- and two-loop effects, both at the origin of 
field space and in the region dominated by leading log potentials.
}

\begin{document}
\maketitle


\section{Introduction}
In the last decade many different mechanisms of supersymmetry breaking have been investigated.
Dynamical supersymmetry breaking (DSB) is an attractive possibility because it can evade constraints imposed by the supertrace formula STr$(\mathcal{M}^2)$. 
Unfortunately, DSB models often lead to non-calculable strongly coupled sectors, in which the knowledge of the spectrum requires 
the use of non-perturbative techniques that are not always available. A new scenario for DSB was proposed in \cite{Intriligator:2006dd}. 
There a weakly coupled IR supersymmetry breaking sector 
was obtained from supersymmetric duality.
A mass deformed  $\mathcal{N}=1$ asymptotically free supersymmetric  
 field theory flows in the IR to a weakly coupled dual theory
with parametrically long-lived metastable minima that break the supersymmetry.
At the  lowest orders in the perturbative expansion the dynamics 
are dominated by a model of pure chiral fields, 
like the O'Raifeartaigh model.
It is therefore important to know the exact and general properties of 
O'Raifeartaigh-like models for the study of DSB.
To provide a phenomenologically viable scenario, we must also break the $R$-symmetry that generically accompanies these models to give the gaugino a non-zero Majorana mass.
 
In this work we focus on $R$-symmetric O'Raifeartaigh-like models whose field content has $R$-charge $R = 0$ or $R = 2$ only.
This property is typical of generalizations of the ISS mechanism
but these models suffer from broad constraints that limit the possibility of spontaneous $R$-symmetry breaking, which is what we seek to achieve in this work.
In particular, the generic (pseudo)moduli fields that accompany the supersymmetry breaking superpotential that have $R = 2$ receive positive corrections from the one-loop Coleman-Weinberg potential, eliminating the possibility of $R$-symmetry breaking via a non-zero modulus field vev, or they remain flat.
There is no general proof for the behavior of the pseudomoduli at higher loops, 
leaving open the possibility of spontaneous $R$-symmetry breaking via higher loop corrections to pseudomoduli that are one-loop flat. 
In fact, most examples that have one-loop flat directions receive negative two-loop corrections that destabilize the origin 
\cite{Giveon:2008wp,Amariti:2008uz}. 
Unfortunately, in all of these examples the tachyonic behavior near the origin is never stabilized at a non-zero pseudomodulus vev, and the potentials run away to a supersymmetric vacuum or infinite field value. 

In principle, a model could be constructed that stabilizes these potentials with tachyonic behavior at the origin by going far out in field space and using the quantum effective potential methods developed by \cite{Intriligator:2008fe}.
It then becomes necessary to introduce one-loop corrections to the pseudomodulus; however, these corrections, at least at the origin, must be subdominant to the tachyonic two-loop effect. We give a rough sketch that shows that having both a tachyonic origin {\em and} a stabilizing ({\em i.e.} positive) slope in the far field potential cannot be accomplished with a single superpotential coupling
if one-loop effects are subdominant at the origin, in the perturbative regime, they will continue to be subdominant to higher-loop order effects far in field space.

This suggests that the myriad obstructions already evident might be evaded by using more than one superpotential coupling. 
A mechanism could be introduced to invert the behavior of the couplings in the two regions of field space and induce the desired behavior of the effective potential. 
More concretely, we invert the natural hierarchy of the perturbative expansion so that at the origin of field space two-loop effects are dominant, but, far in field space, the one-loop effects become more important.
We achieve this through a new coupling associated with massive degrees of freedom that are integrated out at small field values but that contribute far from the origin.
This is reminiscent of the interplay between the gauge and interaction couplings in \cite{Witten:1981kv}, where the coupling hierarchy is inverted in the field space because of asymptotic freedom.

In section \ref{obstructions} we elaborate on the obstructions to spontaneous $R$-symmetry breaking at one and two loops in models with charges $R=0$ and $R=2$ only.
Then in section \ref{RG} we propose our mechanism, explicitly check its validity in a toy model, and
provide a UV completion. In section \ref{conclusions} we conclude and discuss some open questions.

\section{$R$-symmetry breaking with $R=0$ and $R=2$: Obstructions}
\label{obstructions}

The one-loop correction to the mass of the O'Raifeartaigh field is non-negative in models of pure chiral fields with charges $R=0$ and $R=2$ \cite{Shih:2007av}.
This result holds when more than one pseudomodulus is present \cite{Curtin:2012yu}; however, the fate of these pseudomoduli at higher-loop order is generically unconstrained and $R$-symmetry breaking is 
left as a possibility.
Unlike the one-loop Coleman-Weinberg effective potential, 
which can be calculated in terms of the mass matrices only, 
at two-loop order the effective potential must be
explicitly calculated by including the 
Yukawa and quartic couplings\footnote{If the supersymmetry breaking scale $F$ is smaller
  than the messenger scale $M$, $F \ll M^2$, there are simpler results for the two-loop effective potential.
  \cite{Nibbelink:2005wc}}.

Explicit examples show that at two-loop order there are no non-negativity constraints on the pseudomoduli masses as in the one-loop case.
For example, in the model studied in \cite{Giveon:2008wp} the superpotential is\footnote{The model studied in \cite{Giveon:2008wp} is slightly different, but the quantum corrections are computed in a similar manner and the final result is the same.}
\begin{equation} \label{primo}
W = f X + h X \phi_1^2 + h \mu \phi_1 \phi_2 + h  Y \phi_1 \phi_4 + h Z \phi_4^2 + h m \phi_4 \phi_5
\end{equation}
where $\sqrt{f}$, $\mu$, and $m$ are mass scales in the theory, $h$ is the superpotential coupling, the $\phi_i$ fields are tree-level stable at the origin, the $X$ and $Y$ are pseudomoduli stabilized at one-loop, 
and the $Z$ field is still a pseudomodulus at one loop that acquires a negative mass at two loops.

A different possibility has been studied in \cite{Amariti:2008uz}, by starting from the superpotential
\begin{equation} \label{secondo}
W = f X + h X \phi_1^2 + h \mu \phi_1 \phi_2 + h  Y \phi_1 \phi_5 + h Z \phi_4^2 + h m \phi_4 \phi_5.
\end{equation}
In this case the one-loop pseudoflat direction $Z$ has a positive two-loop mass that is stabilized around the 
$R$-symmetric vacuum $\langle Z \rangle =0$.

Clearly, while in (\ref{secondo}) the $R$-symmetry is not spontaneously broken, 
the possibility to break the $R$-symmetry exists in (\ref{primo}).
The vacuum structure for this model must be determined by calculating the behavior of the $Z$ potential away from the origin. 
This can be explored by applying the analysis of \cite{Intriligator:2008fe}.
There, one reconstructs the effective potential for a pseudoflat direction far from the origin but below the cutoff scale by computing the discontinuity in the anomalous dimension of the
massive messengers in the theory.
The pseudomodulus is treated as a background field with non-zero vev. 
By applying this idea the leading log potential is obtained order-by-order in perturbation theory- schematically, with loop order $n$,  one has
\be \label{leadinglog}
V_{eff}(\Phi) \simeq const. + \sum_n (-1)^{(n+1)} \frac{2 }{n!} |f|^2 \Delta{\Omega_X^{(n)}} \log^{n} \frac{|\Phi|}{m_0} 
\ee
and the sign of the coefficient $(-1)^{(n+1)}  \Delta{\Omega_X^{(n)}}$ determines the sign of the potential 
of the pseudomodulus at large $\Phi$.
The discontinuity in the anomalous dimension is captured in $\Delta{\Omega_X^{(n)}} = \left. \frac{d^{n-1}\gamma_X}{dt^{n-1}}\right|^{t_{\Phi}^+}_{t_{\Phi}^-}$.  
As explained in \cite{Intriligator:2008fe}, each derivative of $\gamma_X$ gives a loop factor.
This formula is only valid in the region $\sqrt{F_{\Phi}} \ll \langle \Phi \rangle \ll \Lambda$, where $F_{\Phi}$ is the scale set by the supersymmetry-breaking $F$-terms of $\Phi$. 

In the case of (\ref{primo}) the leading log potential for $Z$ is negative and the potential flows towards a supersymmetric minimum 
(or a runaway)\footnote{The supersymmetric vacuum structure is usually associated with the UV completion of the model.}.
So there are no $R$-symmetry breaking vacua in (\ref{primo}), even though the potential is destabilized at the origin.

One can still try to break the $R$-symmetry with the addition of a tree-level term $W \supset f_2 Z$ to the superpotential. 
Indeed, this term generates a one-loop contribution to the mass of $Z$ (which is automatically positive) and there is a tension between the one- and two-loop contributions, potentially giving a non-supersymmetric vacuum at $\langle Z \rangle \neq 0$.
One can then distinguish the two cases $f_2\simeq f$  and $f_2 \ll f$\footnote{The case $f_2\gg f$ is irrelevant because 
it reverses the role of $Z$ and $X$ in the $f_2 \ll f$ case}.
If $f_2\simeq f$ the positive one-loop correction dominates at the origin and the negative two-loop effect 
dominates at large vev, so the potential has a local maximum at the origin.
On the contrary,  if $f_2 \ll f$ the negative two-loop potential dominates everywhere, since the one-loop effects at the origin are suppressed by $f_2/f \ll 1$. 
In both cases there are no $R$-symmetry breaking minima.

It would appear that this outcome is generic in the models presented. This is argued as follows:
To achieve spontaneous $R$-symmetry breaking in these O'Raifeartaigh-like models, we require that the $Z$ potential be (a) tachyonic at the origin and (b) increasing ({\em i.e.} with positive slope) somewhere further out in field space. 
To satisfy (a), we must have two-loop effects that are dominant at the origin, since one-loop effects will never afford this behavior. 
Since the two-loop effects are suppressed by a factor of $h^2$ compared to the one-loop effects (but aided by a factor of $F_X/F_{\Phi} \gg 1$), this puts a lower bound on the value of $h$\footnote{Hereafter we assume $h$ is a real coupling by absorbing the imaginary part in the phases of the fields.}. 
As we move farther out in field space to the regime where (\ref{leadinglog}) is applicable we begin to lose perturbativity as higher loops become increasingly important.
However, in a model of only chiral fields with one coupling, if the two-loop contribution dominates the one-loop contributions at the origin it will dominate the one-loop contribution everywhere in field space, since no new field content is introduced. 
To satisfy (b), one could argue that the three-loop behavior might accomplish what the one-loop contribution sought to do, but then our ``leading'' log arguments are foregone as we begin to consider all loop contributions. 
More quantitatively, the requirement from (b) that the (positive) one-loop leading log dominate the two-loop leading log out in field space puts an {\em upper} bound on the value of $h$, which will eliminate any parameter space in $h$ left from the previous {\em lower} bound.

We now search for a loophole in this argument based upon the RG properties of the model in the perturbative large field region with multiple couplings. 
In the next section we provide a way to invert the hierarchy amongst the one- and two-loop effects when the potential is dominated by the leading log.

\section{$R$-symmetry breaking from the renormalization group flow}
\label{RG}

We have seen that in O'Raifeartaigh-like models with 
only $R=0$ and $R=2$ fields $R$-symmetry breaking is quite constrained. 
One-loop quantum corrections will leave pseudomoduli flat or stabilize them at the origin, 
while two-loop corrections can be either positive or negative.
At the quantum level, this means that there can exist tension between a positive one-loop and a 
negative two-loop correction\footnote{We ignore higher loop corrections.}.
In the models previously studied this leads to runaway behavior, but here we will attempt to circumvent their fate with a loophole based upon the $RG$ properties of superpotentials and their moduli spaces.

 \subsection{Generalities}
 \label{generalities}
 
Consider a model with chiral fields, a canonical K\"ahler potential and a superpotential $W$ with all fields assigned $R$-charges $R=0$ or $R=2$ such that the $R$-symmetry is preserved.
Let the superpotential be of the form
\begin{equation}
\label{generalmodel}
W = W_1(X,\Phi_i,\phi_i) + W_2 (\Phi_i,\varphi_i)
\end{equation}
where we identify the tree-level flat direction with $X$ and $\Phi_i$
and the other fields are the $\phi_i$ and $\varphi_i$.
The $W_1$ sector has the usual O'Raifeartaigh field $X$ 
in addition to other pseudomoduli $\Phi_i$ and 
massive messengers $\phi_i$. 
The second sector, $W_2$, contains some of the (pseudo)moduli $\Phi_i$ with non-zero $F$-terms such that $F_{\Phi} \ll F_{X}$ and some massive fields $\varphi_i$.  
We assume the masses of the $\varphi_i$ are much larger than those of the $\phi_i$ from the first sector, $m_{\varphi_i} \gg m_{\phi_i}$.

\begin{figure}
\begin{center}
  \includegraphics[width=13cm]{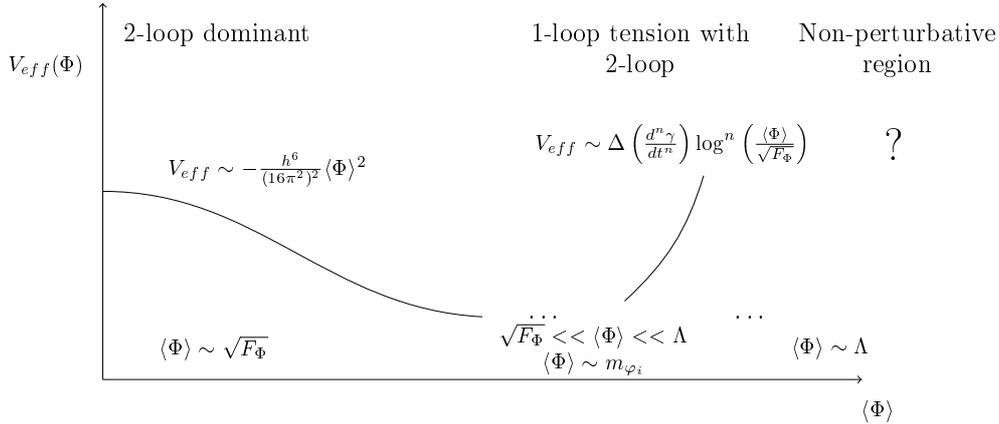}
\caption{A schematic picture of the effective potential for the field 
$\Phi$. Near the origin in $\langle \Phi \rangle$ there is a positive one-loop correction to the tree-level flat potential for $\langle \Phi \rangle$. 
This contribution is suppressed by $\sim \frac{F_{\Phi}}{F_X}$ in comparison to a negative two-loop correction that dominates the one-loop contribution. 
Both are computed perturbatively. 
As we move away from the origin and lose computational control, we approach the far-field region, where $\langle \Phi \rangle \sim m_{\varphi_i}$ but $\langle \Phi \rangle \ll \Lambda$, where $\Lambda$ is the cutoff scale.
Here the potential is computed using the leading log expansion and the one-loop leading $\log \langle \Phi \rangle$ dominates the two-loop leading $\log ^2 \langle \Phi \rangle$ by a careful choice of parameters in the model.
As $\langle \Phi \rangle \sim \Lambda$, we lose all perturbative control over the behavior of the potential.}
\label{figpot}
\end{center}
\end{figure}

In this limit the $W_2$ sector decouples around the origin of the (pseudo)moduli space
and the non-supersymmetric vacuum structure is encrypted in $W_1$\footnote{There is still a non-zero $F$-term associated to $\Phi_i$ in $W_2$ but it is subleading in the limit $F_{\Phi} \ll F_{X}$.}.
We consider $W_1$ such that one of the $\Phi_i$ has a 
vanishing one-loop mass correction but a non-zero, negative two-loop correction.
The effective potential for this field is negative around the origin and it remains negative in the region
$
|F_{X}| \ll |\langle \Phi_i \rangle | \ll \Lambda$, where $\Lambda$ is the strong coupling scale determined by the UV completion of the model \cite{Intriligator:2008fe}.
This model does not generically break the $R$-symmetry spontaneously, at least not without the $W_2$ sector. 
The contributions from $W_2$ become important at a scale $|\langle \Phi_i \rangle | \simeq m_{\varphi_i}$, where the presence of a non-zero $F$-term for $\Phi_i$ gives a positive leading log correction to the effective potential.
The potential in this region is 
\begin{equation}
V(\Phi_i ) \simeq V^{(1)}(h_2, F_{\Phi_i}) + V^{(2)}(h_1, F_{X}) 
\end{equation}
where $h_i$ is the coupling in the $W_i$ sector.
The $R$-symmetry can be broken if $h_2 \gg h_1 \eta(F_X,F_{\Phi_i})$
where $ \eta(F_X,F_{\Phi_i})$ is a model-dependent function. 
Figure \ref{figpot} gives a schematic picture of the effective potential for the field $\Phi$.
\subsection{A toy model}
\label{toy}

Here we propose a toy model that spontaneously breaks the $R$-symmetry
in an O'Raifeartaigh-like model with fields that have $R$-charges $0$ and $2$ only.
We follow the strategy explained above.
The  superpotential $W_1$ is
\begin{equation}\label{w1eqn}
W_1 = f_X X + h_X X \phi_1^2 + m_1 \phi_1 \phi_2 +
Y \phi_1 \phi_4 + h_1 Z \phi_4^2 + m_2 \phi_4 \phi_5 
\end{equation}
while $W_2$ is
\begin{equation}
W_2= f_Z Z 
+h_2 Z \xi_4^2 + m_3 \xi_4 \xi_5
\end{equation}
We impose a hierarchy amongst the scales
\begin{equation} \label{bounds}
\sqrt f_Z \ll \sqrt f_X \ll m_1,m_2 \ll m_3 \ll \Lambda.
\end{equation}

Around the origin, $ \xi_4$ and $ \xi_5$ are integrated out at zero vev and 
the vacuum structure is well described by $W_1$. 
The fields  $\phi_i$ acquire a tree-level mass at zero vev while the fields
$X$ and $Y$
are tree-level flat directions,
 stabilized at the origin by one-loop corrections.
 The field $Z$ is flat at tree level and its quantum mass is
 dominated by the two-loop effect if
\begin{equation}
\epsilon \equiv \frac{f_Z^2}{f_X^2  h_X^2  } \ll 1.
\end{equation}
At larger $\langle Z \rangle$ the effects of $m_3$ are no longer suppressed.
In the region
\begin{equation}
m_3 \ll \langle Z \rangle \ll \Lambda
\end{equation}
the leading log potential is
\begin{equation}
V_{eff} = f_Z^2(h_1^2+h_2^2) \log Z - f_X^2 h_X^2 h_1^2 \log^2 Z .
\end{equation}
There can still be an $R$-symmetry breaking minimum if the inequality 
\begin{equation}
h_2>h_1 \sqrt{\frac{2\log Z}{\epsilon} -1}
 \end{equation}
is satisfied (note this is compatible with (\ref{bounds})).
The $R$-symmetry is broken at the quantum level by the vev of $Z$, with $R(Z) = 2$. 
The presence of two couplings in this simple example follows the construction outlined in section \ref{generalities} and accomplishes spontaneous $R$-symmetry breaking.

   \subsection{A UV completion}
   \label{UVcomp}
  
In this section we discuss a supersymmetric gauge theory
with supersymmetry-breaking metastable vacua that also break its $R$-symmetry.   
In the IR the model reduces to the class introduced above, where $W_1$ and $W_2$ 
provide a generalization of the toy model.
  
\begin{figure}[b]
\begin{center}
\includegraphics[width=10cm]{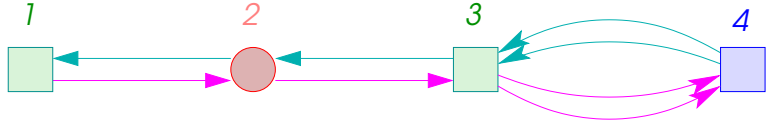}
\caption{A quiver representing the electric theory. The green boxes are flavor nodes, the red the gauge node. We do not fix the 
nature of the blue node: it can be either a flavor symmetry or a weakly gauged global symmetry.}
\label{figele}
\end{center}
\end{figure}

In  this model we tune the masses of the fields  in the UV sector, 
while the tuning on the couplings is dynamical. This 
provides a more natural  explanation of the necessary hierarchies 
amongst the couplings required by our construction.
The field content is (see Figure \ref{figele} for a quiver representation of the model)
\begin{center}
\begin{tabular}{l||cccc}
Field & $SU(N_{F_1})$ & $SU(N_c)$ &$SU(N_{F_2})$ & $SU(M)$ \\
\hline
$Q_{1} \oplus \tilde Q_{1}$&$N_{F_1} + \widetilde{N}_{F_1}$& $\widetilde N_c \oplus N_c$&$1\oplus 1$ &$1\oplus1$\\
$Q_{2} \oplus Q_{2}$&$1\oplus1$&$N_c\oplus \widetilde N_c$&$ \widetilde{N}_{F_2} + {N_{F_2}} $&$1\oplus1$\\
$q_{3} ^{(i)}\oplus \tilde q_{3}^{(i)}$ with $i=1,2$ &$1\oplus1$&$1\oplus1$&$ {N_{F_2}} + \widetilde  {N}_{F_2} $&$\widetilde M\oplus M$ \\
\end{tabular}
\end{center}
with superpotential 
 \begin{equation}\label{uvEqn1}
W = m_1 Q_1 \tilde Q_1 + m_2 Q_2 \tilde Q_2+ m_3 q_3^{(1)} \tilde q_3^{(2)} + m_3 q_3^{(2)} \tilde q_3^{(1)} 
+ \frac{1}{\Lambda_0} Q_2 \tilde Q_2 q_3^{(1)} \tilde q_{3}^{(1)}
\end{equation}
The groups $SU(N_{F_1})$ and $SU(N_{F_2})$ are flavor symmetries while  $SU(N_c)$ 
is the gauge symmetry.  At this level we do not specify the dynamics of $SU(M)$; 
Figure \ref{figele} indicates the possibilities for this $SU(M)$ in the context of a quiver diagram.

\begin{figure}
\begin{center}
\includegraphics[width=10cm]{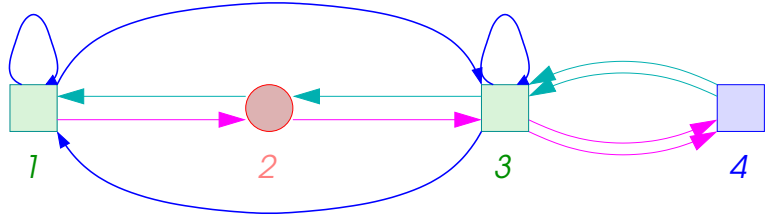}
\caption{A quiver representing the magnetic theory. The green boxes are flavor nodes, the red one is the gauge node, while the 
blue one can be both.}
\label{figmagn}
\end{center}
\end{figure}

We consider this $SU(N_c)$ gauge symmetry in the free magnetic range,
\begin{equation}
N_c+1 < N_{F_1}+N_{F_2} <\frac{3}{2} N_c
\end{equation}
so that the model is described in the IR by the Seiberg dual with field content
(see Figure \ref{figmagn} for the quiver representation)
\begin{center}
\begin{tabular}{l||cccc}
Field & $SU(N_{F_1})$ & $SU(\widetilde N_c)$ &$SU(N_{F_2})$ & $SU(M)$ \\
\hline
$q_{1} \oplus \tilde q_{1}$&$N_{F_1} + \widetilde{N}_{F_1}$& $\widetilde N_c \oplus N_c$&$1\oplus 1$ &$1\oplus1$\\
$q_{2} \oplus q_{2}$&$1\oplus1$&$N_c\oplus \widetilde N_c$&$ \widetilde{N}_{F_2} + {N_{F_2}} $&$1\oplus1$\\
$M_{11}$& $N_{F_1} \times \widetilde{N}_{F_1}$&$1$&$1$&$1$\\
$M_{12} \oplus M_{21}$&$\widetilde N_{F_1} + N_{F_1}$&$1$& $N_{F_2} +\widetilde{N}_{F_2} $&$1$\\
$M_{22}$&$1$&$1$&$N_{F_2} \times \widetilde{N}_{F_2}$&$1$\\
$q_{3} ^{(i)}\oplus \tilde q_{3}^{(i)}$ with $i=1,2$ &$1\oplus1$&$1\oplus1$&$ {N_{F_2}} + \widetilde  {N}_{F_2} $&$\widetilde M\oplus M$ \\
\end{tabular}
\end{center}
where $\widetilde N_c=N_{F_1}+N_{F_2}-N_c$ and the superpotential is
\begin{eqnarray}\label{uvEqn2}
W &=&  h \mu_1^2 M_{11} + h \mu_2^2 M_{22} + 
h 
\left(
\begin{array}{cc}
M_{11}&M_{12}\\
M_{21}&M_{22}
\end{array}
\right)
\left(
\begin{array}{c}
q_1\\
q_2
\end{array}
\right)
\left(
\begin{array}{cc}
\tilde q_1&\tilde q_2
\end{array}
\right) \nonumber 
\\
&+& \frac{\Lambda_g}{\Lambda_0} M_{22} q_3^{(1)} \tilde q_{3}^{(1)}
+ m_3 \left( q_3^{(1)} \tilde q_3^{(2)} + q_3^{(2)} \tilde q_3^{(1)}\right).
\end{eqnarray}
If we fix the hierarchy among the electric masses as 
\begin{equation} \label{regime}
m_3 \gg m_1 \gg m_2
\end{equation}
there is a classical vacuum solution that breaks supersymmetry which can be written as
\begin{eqnarray}
\left(
\begin{array}{c}
q_1\\
q_2
\end{array}
\right) = 
\left(
\begin{array}{c}
  \mu_1 {\mathbf 1}_{\tilde{N}_c \times \tilde{N}_c} \\
  {\mathbf 0}_{(N_{F_1} - \tilde{N}_c) \times \tilde{N}_c} \vspace{-.3cm} \\
  \\
   \hline 
\vspace{-.4cm} \\
{\mathbf 0}_{N_{F_2} \times \tilde{N}_c}
\end{array}
\right) \, , \, 
\left(
\begin{array}{cc}
  \tilde{q}_1 & \tilde{q}_2
\end{array}
\right) = 
\left(
\begin{array}{cccc}
  \mu_1 {\mathbf 1}_{\tilde{N}_c \times \tilde{N}_c} &  {\mathbf 0}_{\tilde{N}_c \times (N_{F_1} - \tilde{N}_c) } &  \vline  & {\mathbf 0}_{\tilde{N}_c \times N_{F_2}}
\end{array}
\right)
\end{eqnarray}
with the rest of the fields at zero expectation value. 
We can expand about this vacuum and choose a convenient parametrization 
of the field fluctuations:
\begin{eqnarray}
\left(
\begin{array}{c}
q_1\\
q_2
\end{array}
\right) = 
\left(
\begin{array}{c}
\mu_1 + \sigma_1\\
\phi_1 \vspace{-.3cm} \\
  \\
   \hline 
\vspace{-.4cm} \\
\phi_5
\end{array}
\right) \, , \, 
\left(
\begin{array}{cc}
  \tilde{q}_1 & \tilde{q}_2
\end{array}
\right) = 
\left(
\begin{array}{cccc}
  \mu_1 + \tilde{\sigma}_1 & \phi_2 & \vline & \phi_4
\end{array}
\right) \nonumber
\end{eqnarray}
\begin{eqnarray}
  M_{11} = 
\left(
\begin{array}{cc}
  \Sigma_{11} & \phi_6\\
\phi_7 & X
\end{array}
\right) \, , \,
  M_{12} = 
\left(
\begin{array}{c}
  \phi_8 \\
  \tilde{Y} 
\end{array}
\right) \, , \,
  M_{21} = 
\left(
\begin{array}{cc}
  \phi_9 &  Y 
\end{array}
\right) \, , \,
M_{22} = Z  \nonumber
\end{eqnarray}
\begin{eqnarray}
\left(
\begin{array}{c}
  q_3^{(1)}\\
  q_3^{(2)}
\end{array}
\right) = 
\left(
\begin{array}{c}
  \xi_4 \\
  \xi_6
\end{array}
\right) \, , \, 
\left(
\begin{array}{cc}
  \tilde{q}_3^{(1)} & \tilde{q}_3^{(2)}
\end{array}
\right) = 
\left(
\begin{array}{cc}
  \xi_5 & \xi_7
\end{array}
\right)
\end{eqnarray}

This yields the IR superpotential 
\begin{eqnarray}\label{irSuperPot}
  W & = & {\rm Tr} [\, h\mu_1^2 X + hX\phi_1\phi_2 + h\mu_1 ( \phi_1\phi_6 + \phi_2\phi_7 + \phi_4\phi_8 + \phi_5\phi_9 ) \nonumber \\
  & + & h\mu_2^2 Z + hZ \phi_4\phi_5 + h_2 Z\xi_4\xi_5 + m_3( \xi_4\xi_6 + \xi_5\xi_7 ) \nonumber \\
  & + & h\phi_1 Y \phi_4 + h\phi_2 \tilde{Y} \phi_5 ]
\end{eqnarray}
plus terms that are supersymmetric at two loops, which is the order to which we study supersymmetry breaking effects in this work. 
Here we have defined $h_2 \equiv \Lambda_g/\Lambda_0$.

From (\ref{regime}) we have
\begin{equation}
m_3^2 \gg h \mu_1^2 \gg h \mu_2^2
\end{equation}
and we can integrate out the $q_3^{(i)}$ and $\tilde q_3^{(i)}$ (with $\langle q_3^{(i)} \rangle =  \langle \tilde q_3^{(i)} \rangle = 0$).
Deep in the IR we have the usual $W_1$ model, with 
$m_Z^2 < 0$ from two-loop quantum effects.
There is also a one-loop contribution that is suppressed ($\mu_1 \gg \mu_2$).
In particular, the one- and two-loop $Z$ masses are
\begin{eqnarray*}
  m_Z^{(1)\, 2} & = \epsilon^4 \frac{h^2\mu_1^2}{24\pi^2}N_{F_2}\tilde{N}_c\left\{ h^2  + h_2^2\left( \frac{\mu_1}{m_3} \right)^2 \right\} + \mathcal{O}\left( \epsilon^6 \right) 
\end{eqnarray*}
\vspace{-1cm}
\begin{eqnarray}\label{loopMasses}
  m_Z^{(2)\, 2} & = \frac{2h^4 \mu_1^2 }{(16\pi ^2)^2}N_{F_2}\tilde{N}_c\left\{ \log 4 -1 -\frac{\pi^2}{6} + \epsilon^4 g(\Lambda)\right\} + \mathcal{O}\left( \epsilon^6 \right) 
\end{eqnarray}
where $g(\Lambda)$ is a complicated but well-behaved function that depends on\footnote{The dependence on the cutoff in $m_Z^{(2)\,2}$ is introduced through the $Z$-self corrections, and vanishes in the limit that $\mu_2 \rightarrow 0$. 
} 
$\log ^2 \Lambda$ and we have defined $\epsilon \equiv \mu_2/\mu_1$.
The trace in (\ref{irSuperPot}) gives a factor of $N_{F_2}\tilde{N}_c$.

\begin{figure}[h]
  \begin{center}
	\includegraphics[width=13cm]{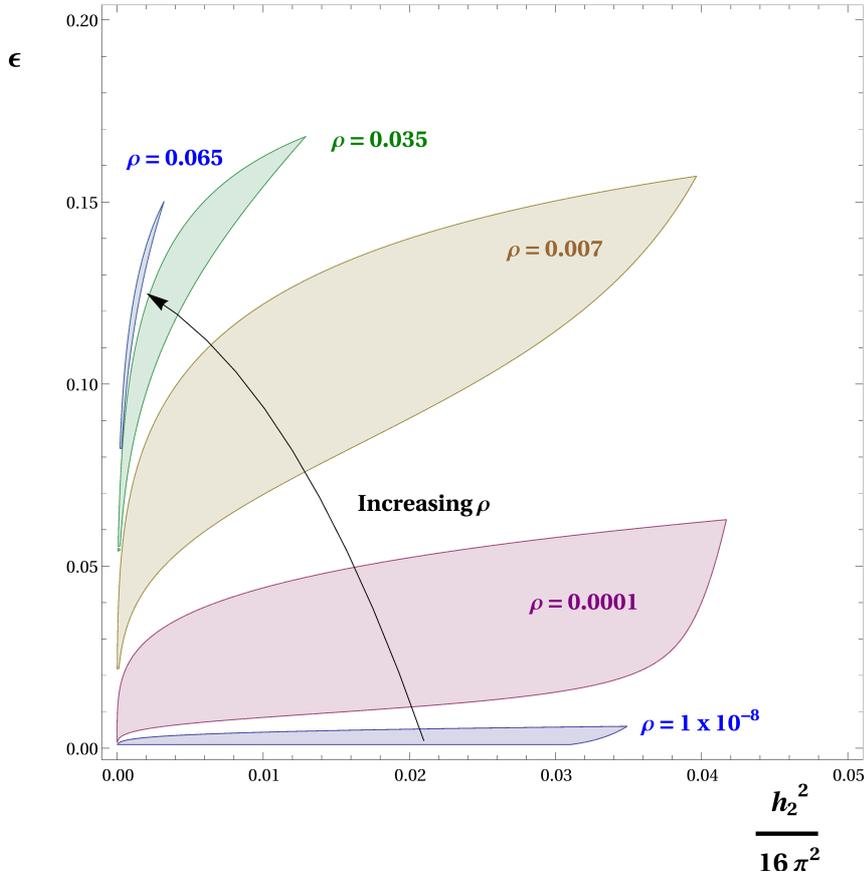}
	\vspace{-1cm}
	\caption{The parameter space that satisfies the requirements (a) that the mass of the $Z$ pseudomodulus is tachyonic at the origin and (b) that the slope of the far field potential be positive. This space is parametrized by the ratio of scales $\epsilon \equiv \frac{\mu_2}{\mu_1}$ and the $m_3$ sector coupling $h_2$ as a function of the ratio $\rho = \frac{h^2}{h_2^2}$. Note that small values of $\rho$ are preferred, but not {\em too} small. The behavior of the allowed regions is smooth everywhere.}
  \label{paramSpace}
  \end{center}
\end{figure}

These masses indicate how the potential for $Z$ behaves at the origin. Clearly, to have a $R$-symmetry breaking minimum, we must have $m_Z^2 = m_Z^{(1)\, 2} + m_Z^{(2)\, 2} < 0$.
However, there is another constraint on the parameters in $m_Z^2$ that comes from the behavior of the potential for $Z$ in the far field region. 
Here $\mu_1 <<  \langle Z \rangle  << \Lambda$; the one- and two-loop contributions to the potential in this region are in tension with one another, since they are introduced with opposite signs ({\em cf}. (\ref{leadinglog})), and we must include the effects of the $m_3$ mass terms.  
Then, to two-loop order, 
\begin{eqnarray}
  V_{ {\rm eff}} (Z)  & = & V^{(1)}_{ {\rm eff}} (Z) + V^{(2)}_{ {\rm eff}} (Z)  \nonumber \\
  & = & \frac{2 \mu_2^4}{16\pi^2}\left( h^2 + h_2^2 \right) \log \frac{\langle Z \rangle}{\mu_2} \nonumber \\
  & \, & \, -  \frac{1}{(16\pi^2)^2}\left( 4\mu_1^4 h^4 \log ^2 \frac{\langle Z \rangle}{\mu_1} + 2\mu_2^4 \left( h^2 + h_2^2 \right)^2 \log ^2 \frac{\langle Z \rangle}{\mu_2} \right)
\end{eqnarray}
up to an unimportant constant. To have a stable $R$-symmetry breaking minimum in the pseudomodulus $Z$, we require that the slope of the potential far in field space be positive, so that an intermediate minimum is guaranteed ({\em cf.} Figure \ref{figpot}).
This further constricts the allowed values of $\epsilon$, $h$, and $h_2$; however, the allowed parameter space is substantial, depending on the ratio between $h$ and $h_2$, which we define as $\rho \equiv \frac{h^2}{h_2^2}$.
Figure \ref{paramSpace} illustrates the allowed values of $\epsilon$ and $h_2$ as a function of the ratio $\rho$.

This model is a \emph{UV completion} of the former toy model in the sense that it provides a gauge theory
that  underlies the model of the chiral fields. 
This completion has two sources of tuning, the first being the mass hierarchy that is necessary to enforce the decoupling of the $W_2$ messenger sector in (\ref{generalmodel}).
There is also tuning in the value of $\rho$. According to Figure \ref{paramSpace}, values where $\rho \ll 1$\footnote{But not {\em too} small, as the two-loop effects in (\ref{w1eqn}) would vanish completely as $\rho \rightarrow 0$! Figure \ref{paramSpace} shows that very small values of $\rho$ are disfavored.} are preferred to maximize the available parameter space. 
For the lowest value depicted, $\rho = 0.0001$, this corresponds to $h \sim \frac{1}{100}  h_2$, but there is still appreciable available parameter space for $h \sim \frac{1}{10} h_2$ $(\rho = 0.01)$.
We also know that this ratio is related to the scales in (\ref{uvEqn1}) and (\ref{uvEqn2}), $\rho = h^2 \left(\frac{\Lambda_0}{\Lambda_{g}} \right)^2$. 
Indeed, dynamically it is more natural to have $\frac{1}{h} \frac{\Lambda_g}{\Lambda_{0}} = \frac{1}{\sqrt{\rho}} \ll 1$ than
the case preferred here, where $\frac{\Lambda_g}{\Lambda_{0}}\gg h$ or $\rho \ll 1$. 
For example, if the quartic term in (\ref{uvEqn1}) arises from a massive field that is integrated out at $\Lambda_0$, then $\Lambda_0$ is roughly its mass and is generically larger than $\Lambda_g$, the duality scale.
  
\begin{figure}[h]
  \begin{center}
	\includegraphics[scale=.3]{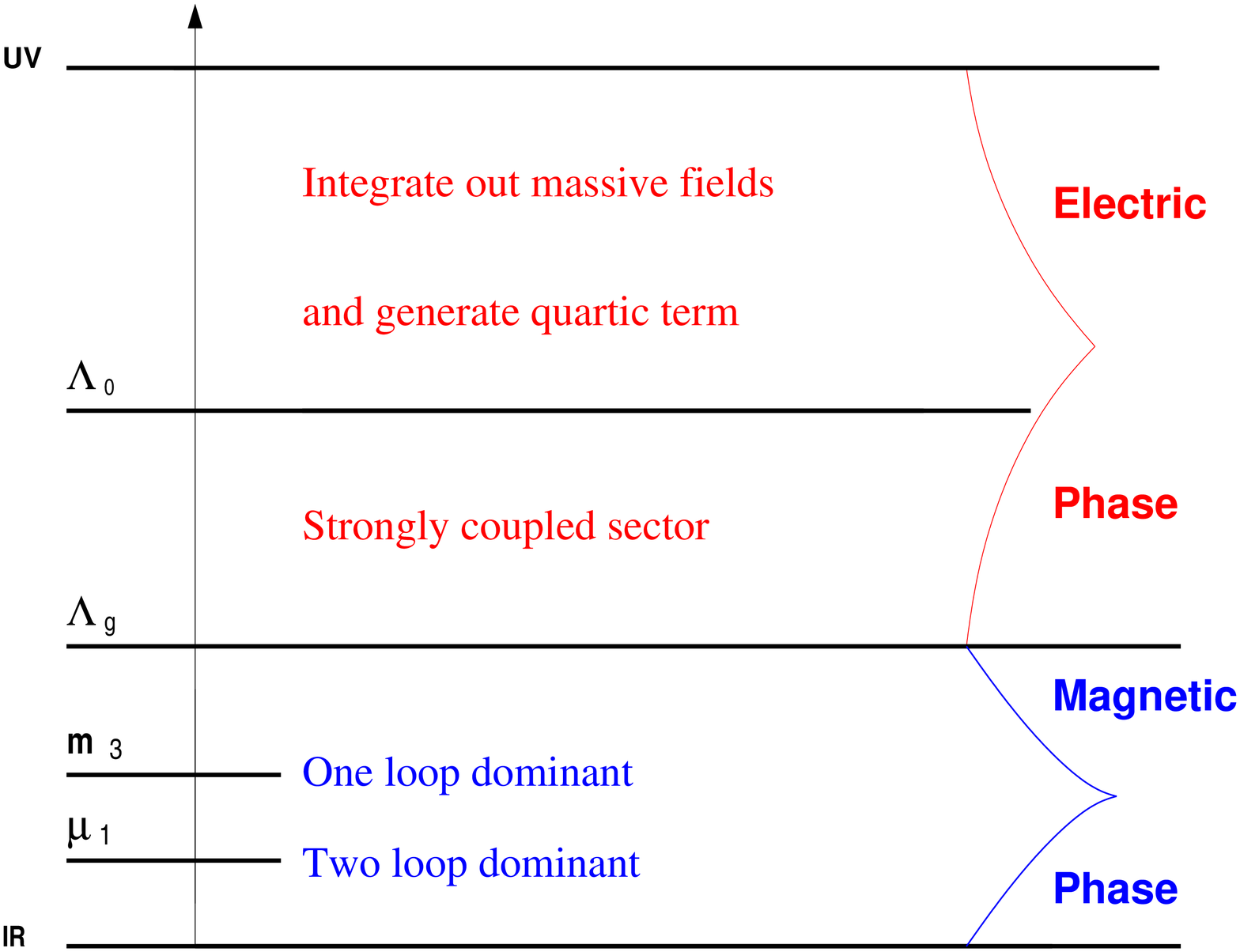}
	\caption{A complete arrangement of the scales introduced into (\ref{irSuperPot}) so as to accomplish spontaneous $R$-symmetry breaking. The ratio $\rho \sim \Lambda_0(\Lambda_g)/\Lambda_g$ is arranged to be smaller than one via running in a strongly coupled sector between the scales $\Lambda_0$ down to $\Lambda_g$. The tuning in $\rho$ is actually quite mild- for appreciable parameter space that allows a spontaneously broken $R$-symmetry, $\rho$ can be as large as $\frac{1}{100}$ (corresponding to $h \sim \frac{1}{10} h_2$- see Figure \ref{paramSpace}).}
	\label{flowPic}
  \end{center}
\end{figure}

The tuning in $\rho$ can be accommodated by assuming that the $h_2$ sector is a generic strongly coupled sector. 
After integrating out the massive field associated to, $\Lambda_0$ the RG flow reduces the effective $\Lambda_0/\Lambda_g$
such that $\frac{\Lambda_g}{\Lambda_{0}(\Lambda_g)}\gg h$ at the scale $\Lambda_g$, where the flow changes. 
These ideas are illustrated in Figure \ref{flowPic}.

\section{Conclusions}
\label{conclusions}

We have shown that there exist $R$-symmetric O'Raifeartaigh-like models with fields having only $R$-charge 0 and 2 that spontaneously break their $R$-symmetry. 
The model we examined had two couplings in the superpotential that exhibited distinct behaviors under their renormalization group flow; in particular, one of the couplings ($h_2$ in (\ref{irSuperPot})) had to be tuned to within $\sim 10$\% by the RG evolution to achieve a spontaneously broken $R$-symmetry.  
There are also two scales in the model that were arranged in a hierarchy, with tuning of order $\sim 10-20$\%.
The $R$-symmetry is broken by the non-zero vev of a $R$-charged pseudomodulus in the model that has a potential dictated by one- and two-loop quantum corrections to its tree-level flat potential. 
The parameter space that allows a non-trivial minimum of the potential is substantial but prefers the tuning in the scales and couplings already mentioned.

Many extensions of our work are possible.
One may look at a brane engineering of the UV model (or some generalizations), as done in \cite{Ooguri:2006bg,Franco:2006ht,
Bena:2006rg} for the ISS model.
It would be interesting to check if the brane action can capture the physics of the non-supersymmetric state that we discovered in this field theory.

Because there is tuning in its marginal couplings, a better understanding of our UV completion is also necessary. A possible explanation of this tuning can come from the strong dynamics of the UV sector- for example, one can suppose that the UV dynamics are governed by an approximate CFT that generates a 
hierarchy amongst the couplings from their anomalous dimensions, as in \cite{Nelson:2000sn}.

One can ponder the possibility of a general result (like in \cite{Shih:2007av}) for the sign of the two-loop masses, possibly associated to some (global) charge assignment. 
In the case of $R=0$ and $R=2$ there is no sign constraint on the mass at two loops, but extra conditions might provide such a constraint (at least at the origin).

We conclude by discussing the embedding of the model in a phenomenological scenario. One can imagine gauging some of the global symmetries and gauge mediating the 
supersymmetry breaking effects to a SSM sector.  This requires the existence of an explicit $R$-symmetry breaking sector to prevent massless axions \cite{Bagger:1994hh}.
It would be important to generate the explicit $R$-symmetry breaking term in the UV theory and study the possible constraints of such a term on the other couplings.

\section*{Acknowledgments}
We are grateful to B. Grinstein, K. Intriligator and A. Mariotti for discussions and comments.
This work was supported by DOE grant DOE-FG03-97ER40546.
\bibliographystyle{JHEP}
\bibliography{References}

\end{document}